\documentclass[letter]{IEEEtran}
\usepackage{amsmath}
\usepackage{amssymb}
\usepackage{amsfonts}
\usepackage{color}
\usepackage[noadjust]{cite}
\usepackage{graphicx} 
\usepackage[utf8]{inputenc}
\usepackage[linesnumbered,ruled]{algorithm2e}
\usepackage{tabularx}

\usepackage{setspace}

\newcommand{\argmax}[1]{\underset{#1}{\operatorname{arg}\,\operatorname{max}}\;}

\makeatletter
\newcommand{\multiline}[1]{%
\begin{tabularx}
	{\dimexpr\linewidth-\ALG@thistlm}[t]{@{}X@{}}#1
\end{tabularx}}

\makeatother

\begin{document}
\title{{IRS-Enhanced OFDMA: Joint Resource Allocation and Passive Beamforming Optimization}
\thanks{The authors are with the Department of Electrical and Computer Engineering, National University of Singapore (email: yifeiyang@u.nus.edu, \{elezhsh,elezhang\}@nus.edu.sg).}}
\author{\IEEEauthorblockN{Yifei~Yang, Shuowen~Zhang,~\IEEEmembership{Member,~IEEE}, and Rui~Zhang,~\IEEEmembership{Fellow,~IEEE}}}
\maketitle
\begin{abstract}
Intelligent reflecting surface (IRS) is an emerging technique to enhance the wireless communication spectral efficiency with low hardware and energy cost. In this letter, we consider the integration of IRS to an orthogonal frequency division multiple access (OFDMA) based multiuser downlink communication system, and study the pertinent joint optimization of the IRS reflection coefficients and OFDMA time-frequency resource block as well as power allocations to maximize the users' common (minimum) rate. Specifically, due to the lack of frequency-selective passive beamforming capability at the IRS, only one set of reflection coefficients can be designed for adapting to a large number of channels of multiple users over different frequency sub-bands. To tackle this difficulty, we propose a novel \emph{dynamic passive beamforming} scheme where the IRS reflection coefficients are dynamically adjusted over different time slots within each channel coherence block to create artificial time-varying channels and select only a subset of the users to be simultaneously served in each time slot, thus achieving a higher passive beamforming gain. Although the formulated optimization problem is non-convex, we propose an efficient algorithm to obtain a suboptimal solution to it. Numerical results show that the proposed scheme significantly improves the system common rate over the setup without IRS and that with random IRS reflection coefficients. Moreover, our proposed dynamic passive beamforming outperforms the fixed passive beamforming which employs a common set of reflection coefficients in each channel coherence block, by more flexibly balancing between passive beamforming and multiuser diversity gains.
\end{abstract}
\vspace{-3mm}
\section{Introduction}
Intelligent reflecting surface (IRS) and its various equivalents have been recently proposed as a new solution to meet the increasingly higher spectral and energy efficiency requirement of future wireless networks \cite{qqmag,ruimag}. By carefully tuning the reflection phase and/or amplitude of a large number of passive elements, IRS is able to proactively alter the wireless propagation environment by creating favorable signal paths between the transmitter and receiver for communication performance enhancement. This is achieved via passive signal reflection only, without the need of any signal generation/amplification, thus incurring much lower energy consumption and hardware cost compared to traditional active beamforming and relaying \cite{qqmag}. Previous studies on IRS-enhanced wireless communication mainly focused on the design of reflection coefficients (or passive beamforming) for narrowband transmission over frequency-flat channels (see, e.g., \cite{qqtwc,debtwc,schober,irssw}), while the more general broadband transmission over frequency-selective channels has been recently studied in \cite{irsglobecom,irstcom,irssw} for the case of {\it single-user} setup. However, the passive beamforming design for \emph{multiuser} broadband communication has not been investigated in the literature, to the best of our knowledge. Compared to the single-user case, the multiuser system design is more involved, as the IRS reflection coefficients need to be jointly optimized with the multiuser transmission scheduling and resource allocation.

In this letter, we consider an IRS-aided multiuser downlink communication system from a base station (BS) to multiple users employing orthogonal frequency division multiple access (OFDMA). We study the joint optimization of the IRS reflection coefficients together with the OFDMA time-frequency resource block (RB) and power allocations in each channel coherence block, for maximizing the minimum (common) rate among all users. To tackle the main difficulty that the IRS reflection coefficients cannot be set different over different frequency sub-channels due to passive reflection, we propose a novel \emph{dynamic passive beamforming} scheme to improve the beamforming performance. Specifically, different IRS reflection coefficients are set for different time slots in each channel coherence block to enable more flexible reflection design with generally less number of users selected to be served at each time slot, thus leading to a higher passive beamforming gain. Although the formulated optimization problem is non-convex and difficult to solve, we propose an efficient algorithm to find a locally optimal solution to it by leveraging alternating optimization and successive convex approximation (SCA) techniques. Numerical results validate the performance gain of our proposed scheme over that without using IRS or with randomly set IRS reflection coefficients, and also show that the proposed dynamic passive beamforming outperforms the fixed passive beamforming where only one common set of IRS reflection \hbox{coefficients is used in each channel coherence block.}
\begin{figure}[t]
	\centering
	\includegraphics[width=5.5cm]{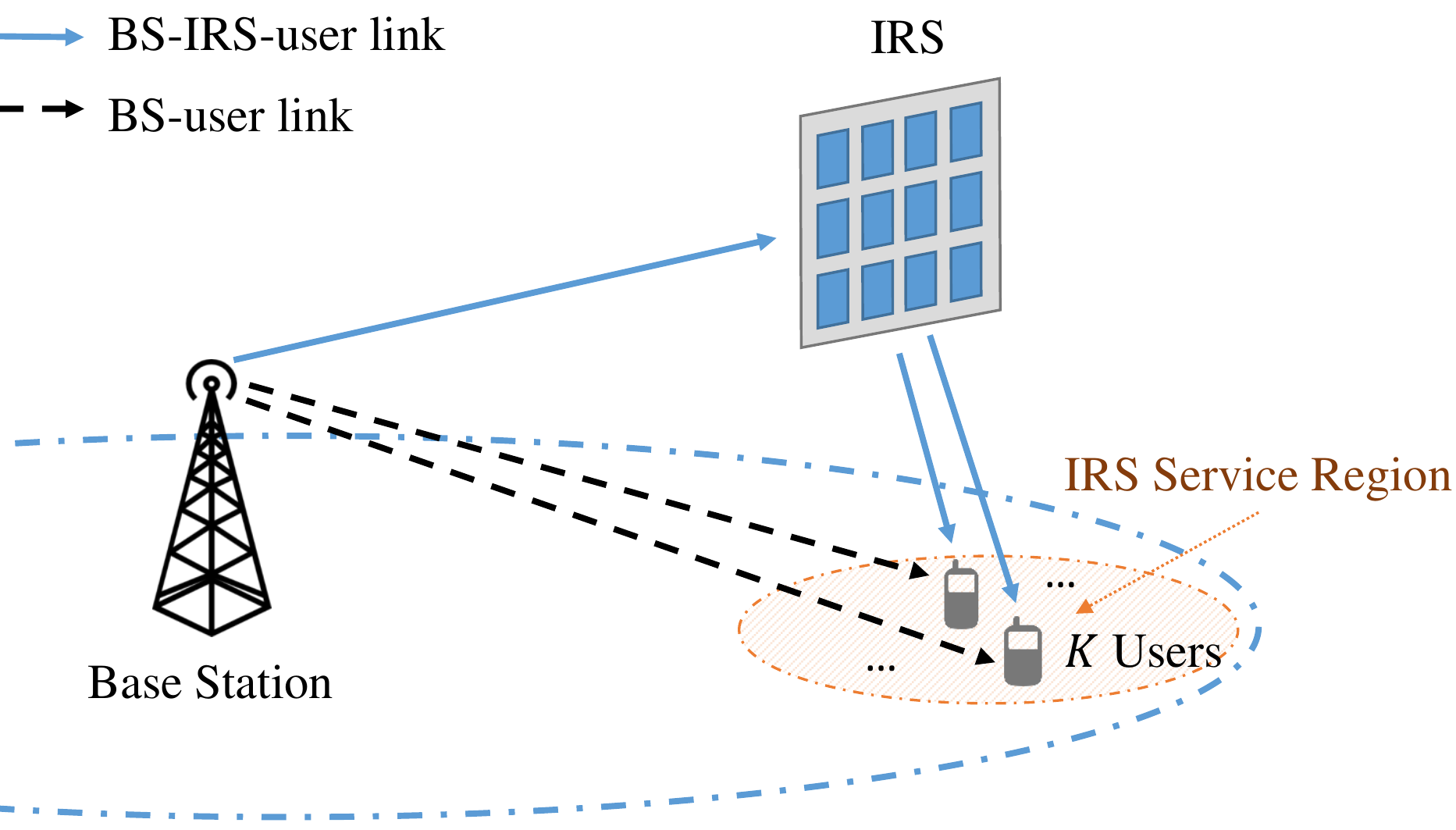}
	\vspace{-2mm}
	\caption{An IRS-enhanced OFDMA downlink communication system.}
	\label{fig:irs}
	\vspace{-6mm}
\end{figure}
\vspace{-5mm}
\section{System Model}\label{sec:sysmodel}
As illustrated in Fig. \ref{fig:irs}, we consider the multiuser downlink communication, where an IRS is deployed to enhance the communications from a single-antenna BS and $K$ single-antenna users located in the vicinity of the IRS, denoted by the set $\mathcal{K}=\{1,\dotsc,K\}$. It is assumed that the IRS consists of $M$ passive reflecting elements, denoted by the set $\mathcal{M}=\{1,\dotsc,M\}$, and is connected to a controller, which adjusts the reflection coefficients of the IRS elements for desired signal reflection. The IRS controller communicates with the BS via a separate control link on the information required for implementing the IRS reflection coefficient design. The signals sent from the BS arrive at each $k$-th user's receiver via two links, namely, the BS-user direct link and the BS-IRS-user reflected link. Let $\tilde{\boldsymbol{h}}^d_k\in\mathbb{C}^{L_{0,k}\times 1}$ denote the baseband equivalent time-domain channel of the BS-user direct link for user $k$, where $L_{0,k}$ denotes the number of delayed taps. Moreover, for each $m$-th reflecting element, let $\tilde{\boldsymbol{t}}_{m}\in\mathbb{C}^{L_1\times 1}$ and $\tilde{\boldsymbol{r}}_{k,m}\in\mathbb{C}^{L_{2,k}\times 1}$ denote the corresponding time-domain BS-IRS channel and that between the IRS and user $k$, respectively, with $L_1$ and $L_{2,k}$ denoting the respective number of delayed taps. Without loss of generality, we consider a quasi-static block fading channel model for all the above channels and focus on one particular channel coherence block where {\hbox{the channels remain approximately constant.}}

\begin{figure}[t]
    \vspace{-1mm}
	\centering
	\includegraphics[width=7cm]{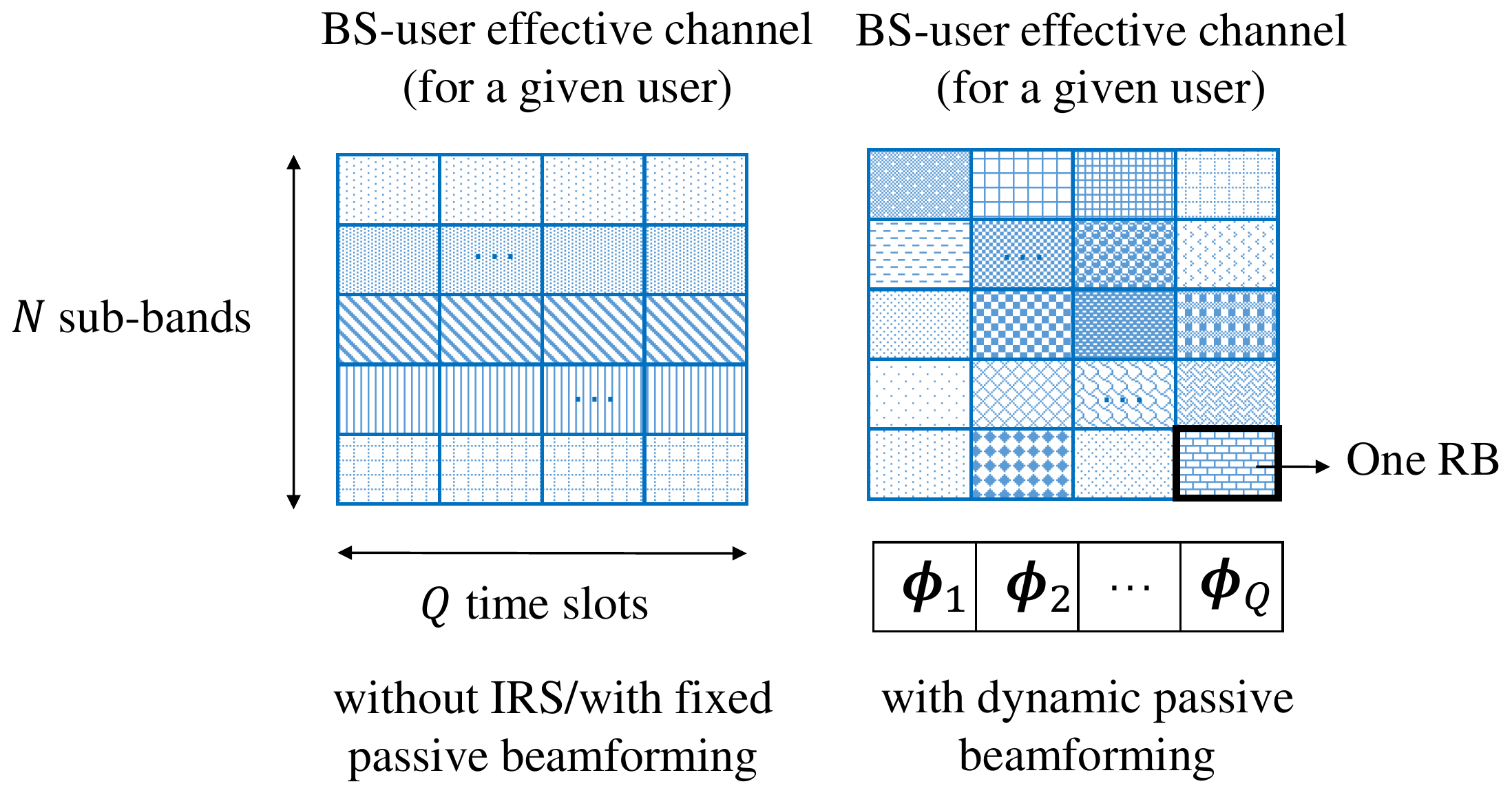}
	\vspace{-3mm}
	\caption{Illustration of RBs and proposed dynamic passive beamforming.}
	\label{fig:rb}
	\vspace{-6mm}
\end{figure}

We consider an OFDMA system where the $K$ users served by the IRS are allocated with $N$ equal sub-bands denoted by the set $\mathcal{N}\!=\!\{1,\dotsc,N\}$, each of which may consist of multiple subcarriers, while the channels are assumed to be constant (frequency-flat) within each sub-band but may vary across different sub-bands. On the other hand, the $K$ users are allocated with $Q$ equal-sized time slots in each channel coherence block, denoted by the set $\mathcal{Q}\!=\!\{1,\dotsc, Q\}$, each of which may consist of multiple orthogonal frequency division modulation (OFDM) symbol durations. In this letter, we consider resource allocation among the $K$ users over the $NQ$ time-frequency RBs, each of which corresponds to one particular pair of sub-band and time slot, as illustrated in Fig. \ref{fig:rb}. It is worth noting that due to the lack of baseband processing capability at the IRS, only a common set of IRS reflection coefficients can be designed for data transmission to the $K$ users over the $N$ sub-bands at any time. This is the main limitation of IRS-aided OFDMA systems as there are in total $NK$ channels that the common reflection coefficients need to cater to in each time slot, which may be prohibitive when $N$ and/or $K$ is large. To overcome this issue, we propose a novel \emph{dynamic passive beamforming} scheme, by allowing the IRS reflection coefficients to be dynamically tuned over different time slots, thereby leading to artificial time-varying channels for each user and consequently more flexibility in the joint resource allocation and reflection coefficient design, as illustrated in Fig. \ref{fig:rb}. In particular, the dynamic passive beamforming is designed such that the corresponding optimal resource allocation allocates generally fewer users to be simultaneously served by the IRS at each time slot, thus reducing the number of channels that the reflection coefficients need to adapt to and thereby enhancing the IRS passive beamforming gain. Note that if the same reflection coefficients are designed over all time slots (named as ``fixed passive beamforming''), then all RBs at each sub-band tend to be allocated to the same user due to their identical channels \hbox{in the same channel coherence block (see Fig. \ref{fig:rb}).}

Specifically, let $\boldsymbol{\phi}_q\!=\![\phi_{q,1},\dotsc,\phi_{q,M}]^T\!\in\!\mathbb{C}^{M\times 1}$ denote the IRS reflection coefficients at each $q$-th time slot with $|\phi_{q,m}|\!\leq \!1$, $\forall q\!\in \!\mathcal{Q}$, $\forall m\!\in\!\mathcal{M}$, due to the passive reflection of the IRS. The effective time-domain reflected channel from the BS to user $k$ via each $m$-th IRS element at each $q$-th time slot is the convolution of the BS-IRS channel, the IRS reflection coefficient, and the IRS-user channel, which is given by $\tilde{\boldsymbol{h}}_{k,m,q}^r\!=\!\tilde{\boldsymbol{r}}_{k,m}*\phi_{q,m}*\tilde{\boldsymbol{t}}_{m}\!=\!\phi_{q,m}\tilde{\boldsymbol{r}}_{k,m}*\tilde{\boldsymbol{t}}_{m}\!\in \!\mathbb{C}^{L_k^r\times 1}$, with $*$ denoting the convolution operation, and $L_k^r\!=\!L_1\!+\!L_{2,k}\!-\!1$ denoting the number of taps in $\tilde{\boldsymbol{h}}_{k,m,q}^r$. For ease of exposition, we further denote for each $k$-th user $\boldsymbol{v}_{k,m}\!=\![(\tilde{\boldsymbol{r}}_{k,m}*\tilde{\boldsymbol{t}}_{m})^T, 0,\dotsc,0]^T\!\in\!\mathbb{C}^{N\times 1}$ as the zero-padded concatenated BS-IRS and IRS-user time-domain channels for each $m$-th reflecting element, and $\boldsymbol{V}_k\!=\![\boldsymbol{v}_{k,1},\dotsc,\boldsymbol{v}_{k,M}]\!\in\!\mathbb{C}^{N\times M}$. The composite BS-IRS-user reflected channel for user $k$ at time slot $q$ can thus be expressed as $\boldsymbol{h}_{k,q}^r\!=\!\boldsymbol{V}_k\boldsymbol{\phi}_q$. By further denoting $\boldsymbol{h}^d_k\!=\![\tilde{\boldsymbol{h}}^d_k,0,\dotsc,0]\!\in\!\mathbb{C}^{N\times 1}$ as the zero-padded time-domain channel for the BS-user direct link, the superposed effective channel \hbox{impulse response (CIR) of user $k$ at time slot $q$ is given by}
\vspace{-1mm}\begin{equation}
{\boldsymbol{h}}_{k,q}=\boldsymbol{h}^d_k+\boldsymbol{h}_{k,q}^r=\boldsymbol{h}^d_k+\boldsymbol{V}_k\boldsymbol{\phi}_q,\quad k\in\mathcal{K}, q\in\mathcal{Q}.
\vspace{-1mm}\end{equation}
Note that there are $L_k=\max(L_{0,k},L_1+L_{2,k}-1)$ non-zero entries in ${\boldsymbol{h}}_{k,q}$, which denotes the number of delayed taps in the effective channel between the BS and user $k$. We further assume that the cyclic prefix (CP) length of the OFDM system is no smaller than the maximum number of delayed taps for the $K$ users, namely, $\underset{k\in \mathcal{K}}{\max}\ L_k$, so that inter-symbol interference (ISI) can be eliminated. The channel frequency response (CFR) at the $n$-th sub-band is then expressed as
\vspace{-1mm}\begin{equation}
c_{k,n,q}=\boldsymbol{f}_n^H\boldsymbol{h}^d_k+\boldsymbol{f}_n^H\boldsymbol{V}_k\boldsymbol{\phi}_q,
\quad k\in\mathcal{K}, n\in\mathcal{N}, q\in\mathcal{Q},
\vspace{-1mm}\end{equation}
where $\boldsymbol{f}_n^H$ denotes the $n$-th row of the $N\times N$ discrete Fourier transform (DFT) matrix $\boldsymbol{F}_N$. For the purpose of exposition, we assume that perfect knowledge of the channels $\boldsymbol{h}^d_k$'s and $\boldsymbol{V}_k$'s is available at the BS, which can be obtained by extending the proposed channel estimation methods in \cite{irstcom,irsbx} to the multiuser case.\footnote{In practice, to reduce the channel training overhead that scales with the number of IRS reflecting elements, we may divide the IRS reflecting elements into multiple groups each consisting of multiple adjacent elements, then only the aggregate channel of each group needs to be estimated \cite{irstcom,irsbx}. It is worth noting that our proposed scheme in this letter is also applicable to this case for low-complexity implementation.} 

Based on the channel state information (CSI), the BS performs sub-band and power allocation as well as dynamic IRS coefficient design, and then informs the IRS and users for communication. Specifically, to avoid inter-user interference, each sub-band at each time slot is allocated to at most one user. Let $\alpha_{k,q,n}$ indicate whether the $n$-th sub-band is allocated to user $k$ at time slot $q$, i.e., $\alpha_{k,q,n}=1$ if sub-band $n$ is assigned to user $k$ at time slot $q$, and $\alpha_{k,q,n}=0$ otherwise.
Thus, we have $\sum_{k=1}^K \alpha_{k,q,n}\leq 1$, $\forall q\in\mathcal{Q}$, $\forall n\in\mathcal{N}$. Moreover, we consider a sum transmit power constraint over the $N$ sub-bands at every time slot $q$, which is given by $\sum_{n=1}^N p_{q,n}\leq P, \;\forall q\in\mathcal{Q}$, where $p_{q,n}\geq 0$ denotes the transmit power allocated to sub-band $n$ at the BS in time slot $q$. Hence, by ignoring the rate loss owing to the CP insertion in each OFDM symbol for ease of exposition, the achievable rate for each $k$-th user in bits per {\hbox{second per Hertz (bps/Hz) is given by}}
\vspace{-2mm}
\begin{equation}
	R_k\!=\!\frac{1}{NQ}\!\sum_{q=1}^Q\!\sum_{n=1}^N \alpha_{k,q,n}\log_2\!\left(\!1\!+\!\frac{|\boldsymbol{f}_n^H\boldsymbol{h}^d_k\!+\!\boldsymbol{f}_n^H\boldsymbol{V}_k\boldsymbol{\phi}_q|^2p_{q,n}}{\Gamma\sigma^2} \right),\label{eqn:rk}
\vspace{-1mm}\end{equation}
where $\Gamma\geq 1$ is the gap from channel capacity due to a practical modulation and coding scheme, and the receiver noise for each sub-band is modeled as an independent and identically distributed (i.i.d.) circularly symmetric complex Gaussian (CSCG) random variable {\hbox{with mean zero and variance $\sigma^2$.}}
\vspace{-4mm}
\section{Problem Formulation}
\vspace{-2mm}
In this letter, we aim to maximize the minimum rate of all users (termed as the ``common rate'' in the sequel) denoted by $R\!=\!\underset{k\in \mathcal{K}}{\min}\ R_k$, by jointly optimizing the IRS dynamic passive beamforming as well as the RB and power allocations at the BS. The optimization problem \hbox{is formulated as}
\begin{subequations}
\begin{align}
\!\!\!\!\!\!\mathrm{(P1)}\!\!\! \underset{\scriptstyle \{\alpha_{k,q,n}\},R\atop \scriptstyle \{{p}_{q,n}\},\{\boldsymbol{\phi}_q\}}{\mathtt{max}} &R	\nonumber\\[-1mm]	
	\mathtt{s.t.}\quad 
	&R_k\geq R,\qquad\qquad \forall k\in\mathcal{K}\label{eqn:constrf}\\[-1mm]
	&\sum_{k=1}^K \alpha_{k,q,n}\leq 1, \quad\  \forall q\in\mathcal{Q},\ \forall n\in\mathcal{N} \label{eqn:constalpha}\\[-1mm]
	&\alpha_{k,q,n}\!\in\!\{0,1\},\ \forall k\in\mathcal{K},\forall q\in\mathcal{Q},\forall n\in\mathcal{N} \label{eqn:constalphakn}\\[-1mm]
	&\sum_{n=1}^{N} p_{q,n}\leq P, \quad\quad \forall q\in\mathcal{Q}\label{P1cd}\\[-1mm]
	&p_{q,n} \geq0, \qquad\quad\quad  \forall q\in\mathcal{Q},\ \forall n\in\mathcal{N}\label{P1ce}\\[-1mm]
	&|\phi_{q,m}|\leq 1,\quad\quad\quad\forall q\in\mathcal{Q},\ \forall m\in\mathcal{M}.\label{eqn:constphikm}
\end{align}
\end{subequations}
(P1) is a non-convex combinatorial optimization problem. In particular, the binary constraints in (\ref{eqn:constalphakn}) are non-convex. Moreover, the achievable rate $R_k$ in (\ref{eqn:constrf}) can be shown to be non-concave over $\boldsymbol{\phi}_q$, while the RB and power allocations as well as the IRS coefficients are all coupled in $R_k$. Therefore, the optimal solution to (P1) is generally difficult to be obtained. In the following section, we propose an efficient algorithm for finding a suboptimal solution to (P1). 
\vspace{-4mm}
\section{Proposed Solution}
\vspace{-2mm}
In this section, we propose an alternating optimization algorithm to find a suboptimal solution to (P1), by iteratively optimizing one of the two sets of optimization variables, namely, the IRS reflection coefficients $\{\boldsymbol{\phi}_q\}$ and the OFDMA resource allocation  $\left\{\{\alpha_{k,q,n}\},\{{p}_{q,n}\}\right\}$, with the other set being fixed at each time, as detailed in the following.
\subsubsection{OFDMA Resource Allocation Optimization Given IRS Coefficients}
With given IRS reflection coefficients $\{\boldsymbol{\phi}_q\}$, the effective channel-to-noise power ratio (CNR) for each $k$-th user at $q$-th time slot and $n$-th sub-band is fixed, which can be obtained as $g_{k,q,n}\overset{\Delta}{=}|\boldsymbol{f}_n^H\boldsymbol{h}^d_k+\boldsymbol{f}_n^H\boldsymbol{V}_k\boldsymbol{\phi}_q|^2/(\Gamma\sigma^2)$, $\forall k\in\mathcal{K}$, $\forall q\in\mathcal{Q}$, $\forall n\in\mathcal{N}$. Problem (P1) thus reduces to the following optimization problem:
\begin{align}
\!\!\mathrm{(P1.1)}
\underset{\scriptstyle \{\alpha_{k,q,n}\}\atop \scriptstyle R,\{p_{q,n}\}}{\mathtt{max}}  &R	\nonumber\\[-3mm]
	\mathtt{s.t.}\quad
	&~\frac{1}{NQ}\!\sum_{q=1}^Q\!\sum_{n=1}^N \alpha_{k,q,n}\log_2\left(1\!+\!g_{k,q,n}p_{q,n}\right)\geq R, \nonumber\\[-2mm]
	&~~~~~~~~~~~~~~~~~~~~~~~~~~~~~~~~~~~ \forall k\in\mathcal{K}\label{eqn:constfd2R}\\[-2mm]
	&~\eqref{eqn:constalpha}-\eqref{P1ce}.\nonumber
\end{align}
Note that (P1.1) is still a non-convex optimization problem due to the binary constraints on $\alpha_{k,q,n}$'s. However, since (P1.1) satisfies the so-called time-sharing condition \cite{weiyu}, the duality gap for (P1.1) is approximately zero for a practically large number of RBs (e.g., $N\!>\!8$ with $Q\!=\!1$ in \cite{weiyu}). Therefore, (P1.1) is similar to traditional OFDMA resource allocation problems (see e.g., \cite{weiyu,cioffiofdma,chengofdma,evansofdma,taoofdma}) with $NQ$ subchannels. Hence, we can adopt the standard \emph{Lagrange duality method} to obtain a high-quality solution to (P1.1) efficiently \cite{weiyu}, for which the complexity can be shown to be $\mathcal{O}((K\!+\!Q)^4\!+\!(K\!+\!Q)^2KNQ)$. Due to the limited space, we omit such details for this algorithm.

\subsubsection{IRS Coefficients Optimization Given OFDMA Resource Allocation}
With given OFDMA resource allocation (i.e., $\{\alpha_{k,q,n}\}$ and $\{p_{q,n}\}$), we aim to optimize a set of IRS reflection coefficients for each time slot (i.e., $\{\boldsymbol{\phi}_q\}$) that maximizes the minimum rate of all users over one channel coherence block. The following problem is thus formulated
\vspace{-1mm}
\begin{align}
\mathrm{(P1.2)}\quad \mathop{\mathtt{max}}_{R,\{\boldsymbol{\phi}_q\}}  &~R	\nonumber\\[-1.5mm]
	\mathtt{s.t.} &~\eqref{eqn:constrf},\eqref{eqn:constphikm}.\nonumber
\end{align}
Note that (P1.2) is still a non-convex optimization problem since the constraints in \eqref{eqn:constrf} can be shown to be non-convex over each $\boldsymbol{\phi}_q$. Hence, in the following, we adopt the SCA technique to obtain a locally optimal solution to (P1.2). Specifically, note that with OFDMA, the transmission over each RB is independent. Thus, by introducing a set of auxiliary variables $\{y_{q,n}\}$, $\{a_{q,n}\}$, and $\{b_{q,n}\}$, (P1.2) can be transformed to the following equivalent problem 
\vspace{-1mm}
\begin{subequations}
\begin{align}
\mathrm{(P1.2')}\qquad &\nonumber\\[-2mm]
\underset{\scriptstyle R,\{\boldsymbol{\phi}_q\},\{y_{q,n}\}\atop \scriptstyle \{a_{q,n}\},\{b_{q,n}\}}{\mathtt{max}}  &~R	\nonumber\\[-5mm]
	\mathtt{s.t.}
	&~\frac{1}{NQ}\sum_{q=1}^Q\sum_{n=1}^N\alpha_{k,q,n}\log_2(1\!+\!\frac{y_{q,n}p_{q,n}}{\Gamma\sigma^2})\geq R,\nonumber\\[-1mm]
	&\qquad \qquad\qquad\qquad\qquad \forall k\in\mathcal{K}\label{eqn:p1f21}\\[-1mm]
	&~|\phi_{q,m}|\!\leq\! 1, \quad\qquad\qquad \forall q\in\mathcal{Q},\forall m\in \mathcal{M}\!\!\\[-1mm]
	&~a_{q,n}\!=\!\Re\left\{\boldsymbol{f}_{n}^H\boldsymbol{h}_{d,\tilde{k}(q,n)}\!+\!\boldsymbol{f}_{n}^H\boldsymbol{V}_{\tilde{k}(q,n)}\boldsymbol{\phi}_q\right\}, \nonumber\\[-1mm]
	&\qquad \qquad\qquad\qquad\qquad \forall q\in\mathcal{Q},\forall n\in\mathcal{N}\\[-1mm]
	&~b_{q,n}\!=\!\Im\left\{\boldsymbol{f}_{n}^H\boldsymbol{h}_{d,\tilde{k}(q,n)}\!+\!\boldsymbol{f}_{n}^H\boldsymbol{V}_{\tilde{k}(q,n)}\boldsymbol{\phi}_q\right\}, \nonumber\\[-1mm]
	&\qquad \qquad\qquad\qquad\qquad \forall q\in\mathcal{Q},\forall n\in\mathcal{N}\label{eqn:p1f22}\\[-1mm]
	&~y_{q,n}\!\leq \!a_{q,n}^2+b_{q,n}^2, \qquad \forall q\in\mathcal{Q},\forall n\in\mathcal{N},\label{eqn:constyn1}
\end{align}
\end{subequations}
where $\tilde{k}(q,n)\!\triangleq\!\{k|\alpha_{k,q,n}\!=\!1, k\!\in\!\mathcal{K}\}$ denotes the user mapping for each RB $(q,n)$, $q\!\in\!\mathcal{Q}$, $n\!\in\!\mathcal{N}$; $\Re\{\cdot\}$ and $\Im\{\cdot\}$ denote the real and imaginary parts of a complex number, respectively. However, (P1.2$'$) is still non-convex due to the non-convex constraints in \eqref{eqn:constyn1}. To tackle this difficulty, we define $\tilde{f}(a_{q,n},b_{q,n})\!\triangleq\! a_{q,n}^2\!+\!b_{q,n}^2$, which is a convex and differentiable function over $a_{q,n}$ and $b_{q,n}$, and lower-bounded by its first-order approximation at $(\tilde{a}_{q,n},\tilde{b}_{q,n})$ as $\tilde{f}(a_{q,n},b_{q,n})\!\geq\!\tilde{a}_{q,n}(2a_{q,n}\!-\!\tilde{a}_{q,n})\!+\!\tilde{b}_{q,n}(2b_{q,n}\!-\!\tilde{b}_{q,n})\!\triangleq\! f(a_{q,n},b_{q,n})$,
where equality holds if and only if $\tilde{a}_{q,n}\!=\!a_{q,n}$ and $\tilde{b}_{q,n}\!=\!b_{q,n}$. Since $f(a_{q,n},b_{q,n})$ is affine over $a_{q,n}$, $b_{q,n}$, and has the same gradient over $(a_{q,n}, b_{q,n})$ as $\tilde{f}(a_{q,n},b_{q,n})$ at $(\tilde{a}_{q,n},\tilde{b}_{q,n})$, we formulate the following problem to solve (P1.2$'$):
\vspace{-2mm}
\begin{align}
\!\!\!\!\!\!\mathrm{(P1.3)}\underset{\scriptstyle R,\{\boldsymbol{\phi}_q\},\{y_{q,n}\}\atop\scriptstyle \{a_{q,n}\},\{b_{q,n}\}}{\mathtt{max}}\!\! &R	\nonumber\\[-1mm]
	\mathtt{s.t.}\ &\eqref{eqn:p1f21}-\eqref{eqn:p1f22}\nonumber\\[-1mm]
	&y_{q,n}\!\leq \!f(a_{q,n},b_{q,n}), \; \forall q\in\mathcal{Q}, \forall n\in\mathcal{N},
\end{align}
which is a convex optimization problem and can be efficiently solved via existing software, e.g., CVX \cite{cvx}, with a complexity of $\mathcal{O}((MNQ)^{4.5}K^{1.5})$ \cite{complexity}. Hence, by adopting the SCA technique, a locally optimal solution to (P1.2) can be obtained by successively updating $\{\tilde{a}_{q,n}\}$ and $\{\tilde{b}_{q,n}\}$ based on the optimal solution to (P1.3) until convergence is achieved \cite{sca}. 

The overall algorithm for solving (P1) is summarized in Algorithm 1, where $I>1$ sets of $\{\boldsymbol{\phi}_q\}$ are randomly generated for initializing the alternating optimization, among which the converged solution with the highest common rate $R$ is selected as the final solution. 
It can be easily verified that the objective function $R$ is non-decreasing over each iteration with a given set of initial $\boldsymbol{\phi}_q$'s, and is upper-bounded by a finite value due to the finite transmit power $P$. Hence, convergence of Algorithm 1 is guaranteed. Moreover, according to \cite{conv}, the obtained solution of Algorithm 1 can be shown to be at least a locally optimal solution to (P1).
\setlength{\textfloatsep}{0pt}
\begin{algorithm}[t]
\DontPrintSemicolon
\caption{Proposed Solution for (P1)}
\KwIn{$\{\boldsymbol{h}_k^d\}$, $\{\boldsymbol{V}_k\}$, $P$, $K$, $N$, $Q$, $\Gamma$, $\sigma^2$, $I$.}
\KwOut{$\{\alpha_{k,q,n}^{\star}\}$, $\{p_{q,n}^{\star}\}$, $\{\boldsymbol{\phi}_q^{\star}\}$, $R^{\star}$.} 
\For{$i=1$ \KwTo $I$}{
Randomly generate $\{\boldsymbol{\phi}_q\}$ with maximum amplitude (equal to one) and uniform phase distribution for each $\phi_{q,m}$, $q\in\mathcal{Q}$, $m\in\mathcal{M}$, as initialization.

\Repeat{\upshape $R$ converges to a prescribed accuracy.}
   	{Fixing $\{\boldsymbol{\phi}_q\}$, optimize $\{\alpha_{k,q,n}\}$ and $\{p_{q,n}\}$ via Lagrangian duality method.\\[-0.5mm]
   	
	\hbox{Fixing $\{\alpha_{k,q,n}\!\}$ and $\{p_{q,n}\!\}$, update $\{\boldsymbol{\phi}_q\!\}$ via SCA.}
}

Record the optimal solution as $\left(\!\{\alpha_{k,q,n}^{(i)}\}, \{p_{q,n}^{(i)}\}, \{\boldsymbol{\phi}_q^{(i)}\}, R^{(i)}\!\right)$.
}

Set $i^{\star}=\argmax{i=1,\dotsc, I} R^{(i)}$, $\left(\{\alpha_{k,q,n}^{\star}\},\{p_{q,n}^{\star}\},\{\boldsymbol{\phi}_q^{\star}\} \right)\leftarrow\left(\{\alpha_{k,q,n}^{(i^{\star})}\},\{p_{q,n}^{(i^{\star})}\},\{\boldsymbol{\phi}_q^{(i^{\star})}\} \right)$
\end{algorithm}

It is worth noting that unlike traditional OFDMA where the channels for all users remain static throughout each channel coherence block, by properly designing the IRS reflection coefficients over different time slots within each channel coherence block, we are able to proactively generate optimized artificial time-varying channels, which help enhance the multiuser diversity over time and allow for more flexible OFDMA resource allocation, as will be illustrated in the next section.

\vspace{-4mm}
\section{Numerical Results}\label{sec:sim}
\begin{figure}[!b]
    \centering
    \includegraphics[width=0.55\linewidth, keepaspectratio]{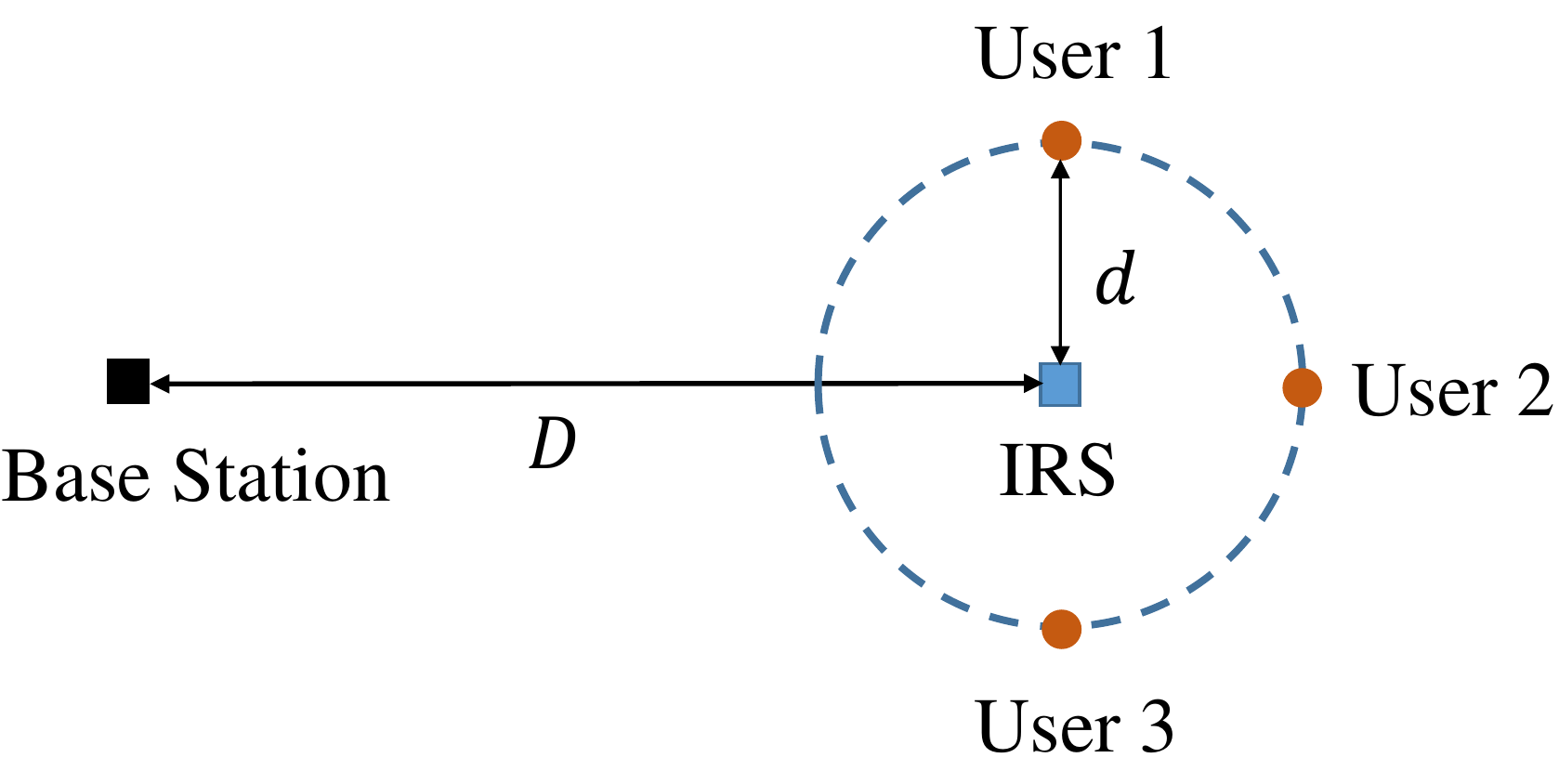}
    \vspace{-3mm}
    \caption{Locations of the BS, IRS, and users for simulations.}
    \label{fig:irs_sim}
\end{figure}

\vspace{-1mm}
In this section, we evaluate the performance of the proposed IRS-enhanced OFDMA system. For the purpose of exposition, we consider a coherence block with $Q=6$ time slots allocated to IRS-aided users over $N=16$ sub-bands. For illustration, we consider the system setup shown in Fig. \ref{fig:irs_sim} with $K=3$ users, where the users are located on a semi-circle around the IRS with distance $d$, and we set $D=100$ meters (m) and $d=2$ m.
The path loss of each channel is modeled by 
$\zeta\!=\!\zeta_0\left(D'/D_0\right)^{-\beta}$,
where $\zeta_0\!=\!-30$ dB denotes the reference path loss at the reference distance $D_0\!=\!1$ m, $D'$ denotes the link distance, and $\beta$ denotes the path loss exponent. The path loss exponent for the BS-user channel, the BS-IRS channel, and the IRS-user channel is set as $\beta_{Bu}\!=\!3.5$, $\beta_{BI}\!=\!2.2$, and $\beta_{Iu}\!=\!2.8$, respectively. For each multi-path channel defined in Section \ref{sec:sysmodel}, the channel taps are given by $g_l\!=\!\sqrt{\zeta\frac{t_{l}}{\sum_{l'=0}^{L-1}t_{l'}}}\xi,\ l\!=\!0,\dotsc,L-1$, where $\zeta$ is the path loss of the link modeled above, $t_l/(\sum_{l'=0}^{L-1}t_{l'})$ is the normalized tap power of the $l$-th tap, which is assumed to follow an exponential power delay profile (i.e., $t_l=e^{-l/(L-1)}$), $\xi$ is modeled as an i.i.d. CSCG random variable with mean zero and variance one (i.e., $\xi\sim\mathcal{CN}(0,1)$). The maximum tap delay for the BS-user direct link, BS-IRS link, and IRS-user link is set as $L_{0,k}\!=\!4,\forall k\!\in\! \mathcal{K}$, $L_{1}\!=\!2$, and $L_{2,k}\!=\!3,\forall k\!\in \!\mathcal{K}$, respectively. The total transmission power at the BS for each time slot is set as $P\!=\!35$ dBm, whereas the receiver noise at each sub-band is set as $\sigma^2\!=\!-110$ dBm. The number of randomizations for the initialization of Algorithm 1 is set as $I\!=\!5$, and the SNR gap is set as $\Gamma=8.8$ dB. All simulations are averaged over 100 independent channel realizations.

Besides the proposed joint resource allocation and dynamic passive beamforming design, we also consider three benchmark schemes as follows.
\begin{enumerate}
	\item \textbf{Fixed Passive Beamforming}: We consider that one set of IRS reflection coefficients is used throughout the $Q$ time slots, i.e., $\boldsymbol{\phi}_q=\boldsymbol{\phi},\forall q\in \mathcal{Q}$. The OFDMA resource allocation and the IRS reflection coefficients are then optimized similarly as the dynamic passive beamforming case via Algorithms 1.
	\item \textbf{Random Phase}: The IRS reflection coefficients for the entire coherence block (Random Phase I) or at each time slot (Random Phase II) are generated with random phase shifts (uniformly distributed within $[-\pi,\pi)$) and maximum reflection amplitude (i.e., unity), based on which the OFDMA resource allocation is optimized. 
	\item \textbf{Without IRS}: We only consider the BS-user direct link and perform the OFDMA resource allocation. 
\end{enumerate}

\begin{figure}[!b]
    \centering
    \includegraphics[width=8cm]{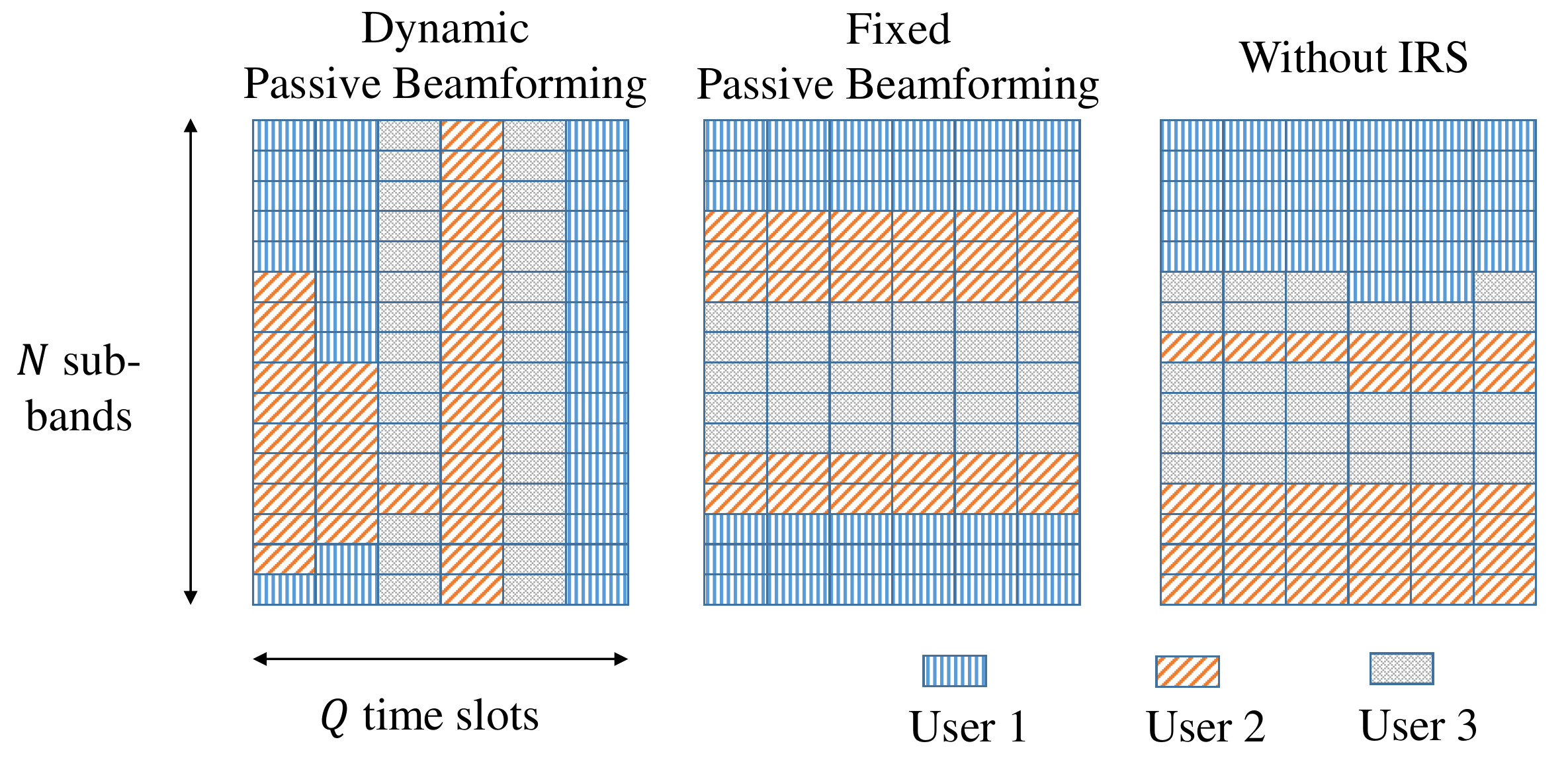}
    \vspace{-3mm}
    \caption{Illustration of the optimal OFDMA RB allocation with $M=80$.}
    \label{fig:alloc1}
\end{figure}
First, we investigate the OFDMA RB allocation pattern for different schemes. Fig. \ref{fig:alloc1} shows the obtained optimal RB allocation using the proposed and benchmark algorithms for a random channel realization with $M=80$. It is worth noting that with fixed passive beamforming or without IRS, the RB allocation is almost invariant over the time slots. This is because in these two cases, the CFR is invariant over the $Q$ time slots for each user, as shown in Fig. \ref{fig:rb}. In contrast, with dynamic passive beamforming, a distinct RB allocation pattern is observed. Specifically, the RBs in each time slot are allocated to fewer users, or even to a single user in some time slots. This is due to that the IRS reflection coefficients are not frequency-selective, as a result it is generally difficult to find one set of coefficients for catering to all channels of all users at each time slot. Hence, by scheduling fewer users in one time slot, higher beamforming gain can be obtained as the coefficients are customized for fewer channels. Although this leads to reduced multiuser diversity gain at each time slot, it is worth noting that the passive beamforming gain generally increases with $M$, while the multiuser diversity gain is invariant with $M$. Thus, the passive beamforming gain overwhelms the multiuser diversity gain at large $M$ and it is expected that the optimal RB allocation approaches that of time division multiple access (TDMA) at large $M$ and $Q$, i.e., each time slot is allocated with one user only.
\begin{figure}[!t]
    \centering
    \includegraphics[width=7.5cm]{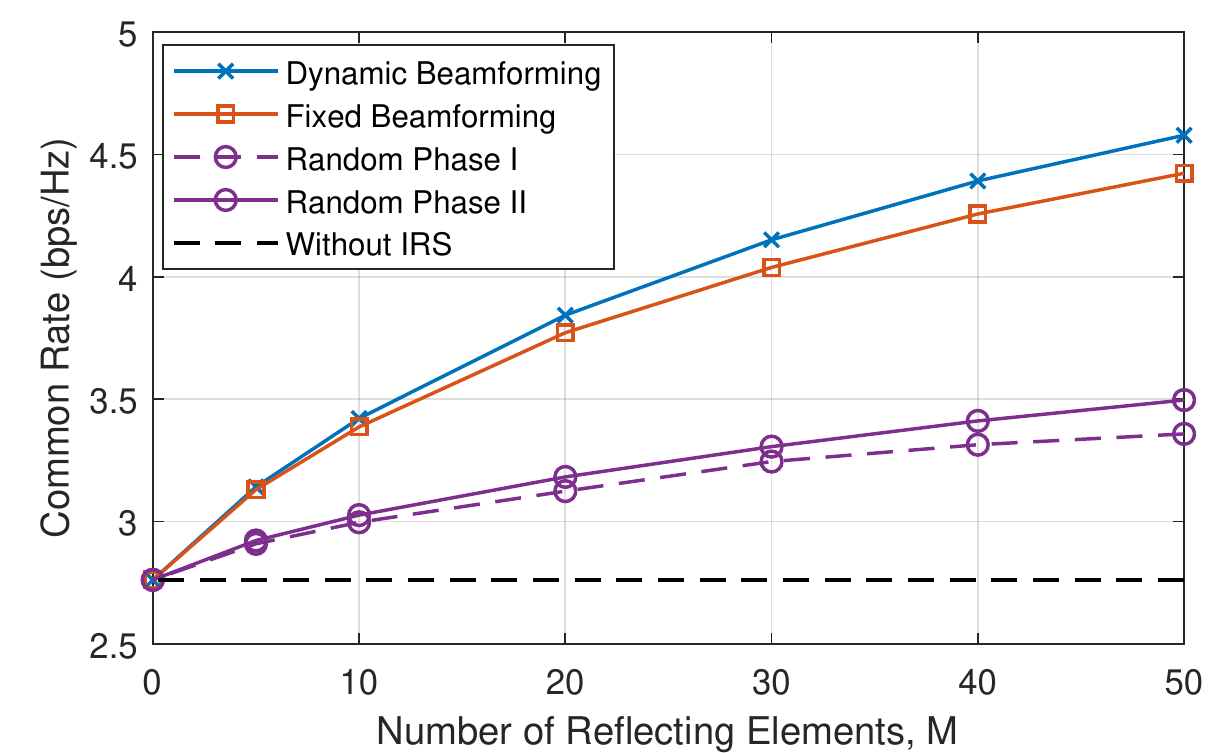}
    \vspace{-3mm}
    \caption{Multiuser common rate versus $M$.}
    \label{fig:M}
\end{figure}

Fig. \ref{fig:M} shows the multiuser common rate versus the number of reflecting elements $M$. It is observed that as $M$ increases, all the schemes with IRS outperform the case without IRS. Moreover, both dynamic passive beamforming and fixed passive beamforming outperform the cases with random phase and the performance gap increases with $M$, which shows the gain of channel-based passive beamforming designs in IRS-enhanced OFDMA systems. Meanwhile, it is observed that Random Phase II outperforms Random Phase I due to the enhanced multiuser diversity; while the scheme with dynamic passive beamforming yields higher common rate compared to that with fixed passive beamforming, since our proposed scheme more flexibly balances between the multiuser diversity gain and the IRS \hbox{passive beamforming gain.}
\vspace{-3mm}
\section{Conclusion}
In this letter, we considered an IRS-enhanced OFDMA system and investigated the joint optimization of IRS passive beamforming and OFDMA resource allocation to maximize the multiuser common rate. A novel dynamic passive beamforming scheme was proposed to induce artificial channel fading and thereby improve the passive beamforming performance. An efficient alternating optimization algorithm was proposed to find a locally optimal solution for the formulated problem. Effectiveness of the proposed algorithm was validated via numerical results, which showed that deploying the IRS greatly improves the multiuser common rate. It was also shown that with dynamic passive beamforming, the OFDMA RB allocation becomes more dynamic as compared to that without IRS or with fixed passive beamforming, with fewer users simultaneously served in each time slot. This leads to improved common rate by more flexibly balancing between the multiuser diversity and IRS passive beamforming gains.
\vspace{-3mm}
\begin{spacing}{0.86}
	\bibliographystyle{IEEEtran}
	\bibliography{irs}

\begin{thebibliography}{10}
\providecommand{\url}[1]{#1}
\csname url@samestyle\endcsname
\providecommand{\newblock}{\relax}
\providecommand{\bibinfo}[2]{#2}
\providecommand{\BIBentrySTDinterwordspacing}{\spaceskip=0pt\relax}
\providecommand{\BIBentryALTinterwordstretchfactor}{4}
\providecommand{\BIBentryALTinterwordspacing}{\spaceskip=\fontdimen2\font plus
\BIBentryALTinterwordstretchfactor\fontdimen3\font minus
  \fontdimen4\font\relax}
\providecommand{\BIBforeignlanguage}[2]{{%
\expandafter\ifx\csname l@#1\endcsname\relax
\typeout{** WARNING: IEEEtran.bst: No hyphenation pattern has been}%
\typeout{** loaded for the language `#1'. Using the pattern for}%
\typeout{** the default language instead.}%
\else
\language=\csname l@#1\endcsname
\fi
#2}}
\providecommand{\BIBdecl}{\relax}
\BIBdecl

\bibitem{qqmag}
Q.~Wu and R.~Zhang, ``Towards smart and reconfigurable environment: Intelligent
  reflecting surface aided wireless network,'' to appear in \textit{IEEE
  Commun. Mag.} [Online]. Available: https://arxiv.org/abs/1905.00152.

\bibitem{ruimag}
E.~Basar, M.~D. Renzo, J.~D. Rosny, M.~Debbah, M.~S. Alouini, and R.~Zhang,
  ``Wireless communications through reconfigurable intelligent surfaces,''
  \emph{IEEE Access.}, vol.~7, pp. 116\,753--117\,773, Sept. 2019.

\bibitem{qqtwc}
Q.~Wu and R.~Zhang, ``Intelligent reflecting surface enhanced wireless network
  via joint active and passive beamforming,'' \emph{IEEE Trans. Wireless
  Commun.}, vol.~18, no.~11, pp. 5394--5409, Nov. 2019.

\bibitem{debtwc}
C.~Huang, A.~Zappone, G.~C. Alexandropoulos, M.~Debbah, and C.~Yuen,
  ``Reconfigurable intelligent surfaces for energy efficiency in wireless
  communication,'' \emph{IEEE Trans. Wireless Commun.}, vol.~18, no.~8, pp.
  4157--4170, Aug. 2019.

\bibitem{schober}
X.~Yu, D.~Xu, and R.~Schober, ``{MISO} wireless communication systems via
  intelligent reflecting surfaces,'' in \emph{Proc. IEEE Int. Conf. Commun.
  China (ICCC)}, Changchun, China, Aug. 2019, pp. 735--740.

\bibitem{irssw}
S.~Zhang and R.~Zhang, ``Capacity characterization for intelligent reflecting
  surface aided {MIMO} communication,'' [Online]. Available:
  https://arxiv.org/abs/1910.01573.

\bibitem{irsglobecom}
Y.~Yang, S.~Zhang, and R.~Zhang, ``{IRS}-enhanced {OFDM}: Power allocation and
  passive array optimization,'' to appear in \textit{Proc. IEEE Global Commun.
  Conf. (Globecom)}, Waikoloa, HI, 2019 [Online]. Available:
  https://arxiv.org/abs/1905.00604.

\bibitem{irstcom}
Y.~Yang, B.~Zheng, S.~Zhang, and R.~Zhang, ``Intelligent reflecting surface
  meets {OFDM}: Protocol design and rate maximization,'' [Online]. Available:
  https://arxiv.org/pdf/1906.09956.

\bibitem{irsbx}
B.~Zheng and R.~Zhang, ``Intelligent reflecting surface enhanced
  \protect{OFDM}: channel estimation and reflection optimization,'' [Online].
  Available: https://arxiv.org/abs/1909.03272.

\bibitem{weiyu}
W.~Yu and R.~Lui, ``Dual methods for nonconvex spectrum optimization of
  multicarrier systems,'' \emph{IEEE Trans. Commun.}, vol.~54, no.~7, pp.
  1310--1322, Jul. 2006.

\bibitem{cioffiofdma}
W.~Rhee and J.~M. Cioffi, ``Increase in capacity of multiuser {OFDM} system
  using dynamic subchannel allocation,'' in \emph{Proc. IEEE Veh. Technol.
  Conf. (VTC Spring)}, Tokyo, Japan, May 2000, pp. 1085--1089.

\bibitem{chengofdma}
Y.~M. Tsang and R.~S. Cheng, ``Optimal resource allocation in
  {SDMA}/{Multi-Input-Single-Output}/{OFDM} systems under {QoS} and power
  constraints,'' in \emph{Proc. IEEE Wireless Commun. Network. Conf. (WCNC)},
  Atlanta, GA, Mar. 2004, pp. 1595--1600.

\bibitem{evansofdma}
Z.~Shen, J.~G. Andrews, and B.~L. Evans, ``Adaptive resource allocation in
  multiuser {OFDM} systems with proportional rate constraints,'' \emph{IEEE
  Trans. Wireless Commun.}, vol.~4, no.~6, pp. 2726--2737, Nov. 2005.

\bibitem{taoofdma}
M.~Tao, Y.~C. Liang, and F.~Zhang, ``Resource allocation for delay
  differentiated traffic in multiuser {OFDM} systems,'' \emph{IEEE Trans.
  Wireless Commun.}, vol.~7, no.~6, pp. 2190--2201, Jun. 2008.

\bibitem{cvx}
M.~Grant and S.~Boyd, ``\protect{CVX: Matlab Software for Disciplined Convex
  Programming},'' \protect{Dec}. 2018 [Online]. Available:
  http://cvxr.com/cvx/.

\bibitem{complexity}
K.~Y. Wang, A.~M.~C. So, T.~H. Chang, W.~K. Ma, and C.~Y. Chi, ``Outage
  constrained robust transmit optimization for multiuser {MISO} downlinks:
  Tractable approximations by conic optimization,'' \emph{IEEE Trans. Signal
  Process.}, vol.~62, no.~21, pp. 5690--5705, Nov. 2014.

\bibitem{sca}
B.~R. Marks and G.~P. Wright, ``A general inner approximation algorithm for
  nonconvex mathematical programs,'' \emph{Operations Research}, vol.~26,
  no.~4, pp. 681--683, Jul. 1978.

\bibitem{conv}
M.~{Hong}, M.~{Razaviyayn}, Z.~{Luo}, and J.~{Pang}, ``A unified algorithmic
  framework for block-structured optimization involving big data: With
  applications in machine learning and signal processing,'' \emph{IEEE Signal
  Process. Mag.}, vol.~33, no.~1, pp. 57--77, Jan. 2016.

\end{thebibliography}
\end{spacing}

\end{document}